\documentclass[12pt]{article}
\usepackage{epsfig}
\usepackage{amsfonts}
\usepackage{amssymb}

\parskip=\medskipamount	
\renewcommand{\thesection}{\arabic{section}}

\catcode`@=11
\@addtoreset{equation}{section}
\@addtoreset{equation}{subsection}
\def\theequation{\ifnum\value{section}=0 \arabic{equation}\ignorespaces
\else \ifnum\value{section}=-1 A.\arabic{equation}\ignorespaces
\else \ifnum\value{subsection}=0 \thesection.\arabic{equation}\ignorespaces
\else \thesection.\arabic{subsection}.\arabic{equation}\ignorespaces
                              \fi
                         \fi
                    \fi}
\catcode`@=12

\newcommand{\bq}{\begin{equation}}
\newcommand{\fq}{\end{equation}}
\newcommand{\bqr}{\begin{eqnarray}}
\newcommand{\fqr}{\end{eqnarray}}

\newcommand{\rf}[1]{(\ref{#1})}

\def\alp{\alpha}       
   \def\eps{\epsilon}

\def\cA{{\cal A}}  
  
  \def\cI{{\cal I}}
  
 \def\cN{{\cal N}}

\def\pa{\partial}

\def\sfrac#1#2{{\textstyle{\frac#1#2}}}

\def\ab{[{\textstyle{\alp\atop\beta}}]}

\def\on#1#2{{\buildrel{\mkern2.5mu#1\mkern-2.5mu}\over{#2}}}
\def\ron#1#2{{\buildrel{#1}\over{#2}}}	
\def\dt#1{\on{\hbox{\bf .}}{#1}}                
\def\rdt#1{\ron{\hbox{\bf .}}{#1}}

\def\da{\dt\alpha}
\def\db{\dt\beta}
\def\pd{{\rdt{+}}}
\def\md{{\rdt{-}}}

\def\bop#1{\setbox0=\hbox{$#1M$}\mkern1.5mu
	\vbox{\hrule height0pt depth.04\ht0
	\hbox{\vrule width.04\ht0 height.9\ht0 \kern.9\ht0
	\vrule width.04\ht0}\hrule height.04\ht0}\mkern1.5mu}
\def\Box{{\mathpalette\bop{}}}                        

\begin{document}

\thispagestyle{empty}

\begin{flushright}
\begin{tabular}{l}
ANL-HEP-PR-00-086 \\
YITP-SB-00-54 \\
hep-th/0010238
\end{tabular}
\end{flushright}

\vskip.5in minus.2in
\begin{center}

{\Large\bf Global Conformal Anomaly in $\cN$ = 2 String}\\

\vskip.5in minus.2in

{\bf Gordon Chalmers}
\\[5mm]
{\em Argonne National Laboratory \\
High Energy Physics Division \\
9700 South Cass Avenue \\
Argonne, IL 60439-4815 } \\
{Email: chalmers@pcl9.hep.anl.gov}\\[5mm]

\vskip .2in
{\bf Warren Siegel}
\\[5mm]
{\em C.N. Yang Institute for Theoretical Physics \\
State University of New York at Stony Brook \\
Stony Brook, NY 11794-3840}\\
{Email: siegel@insti.physics.sunysb.edu}

\vskip.5in minus.2in

{\bf Abstract}

\end{center}

We show the existence of a global anomaly in the one-loop graphs of
$\cN=2$ string theory, defined by sewing tree amplitudes, unless
spacetime supersymmetry is imposed.  The anomaly is responsible for the
non-vanishing maximally helicity violating amplitudes.
The supersymmetric completion of the $\cN=2$ string spectrum is
formulated by extending the  previous cohomological analysis with an
external spin factor; the target space-time spin-statistics of these
individual fields in a selfdual background are compatible with previous
cohomological analysis as fields of arbitrary spin may be bosonized into
one another.  We further analyze duality relations between the open and
closed string amplitudes and  demonstrate this in the supersymmetric
extension of the target space-time theory through the insertion of
zero-momentum operators.

\setcounter{page}{0}
\newpage
\setcounter{footnote}{0}


\section{Introduction}

The $\cN=2$ string is unique among string theories, as its critical
dimension is four and it contains only massless states in its spectrum. It
has two local supersymmetries on the world-sheet, and its local
world-sheet action consists of the $\cN=2$ gravitational
multiplet coupled to two complex chiral supermultiplets, and was
originally proposed and examined in \cite{Ademollo:1976an}. This
string theory has topological space-time properties intrinsic to the
world-sheet, linked to a vector in the gravitational multiplet.

\subsection{Classical $\cN=2$ strings}

Ooguri and Vafa \cite{Ooguri:1991fp} (see also Marcus
\cite{Marcus:1992wi}), in  the correct critical dimension of four
(space-time $4+0$ or $2+2$), found the classical  theory to describe in the
target space-time classical selfdual gravity  in the closed $\cN=2$ string
and classical selfdual Yang-Mills theory in the open  formulation. The
target space-time light-cone action $S = {\rm Tr} \int d^4 x\ {\cal 
L}$ for the closed  string is given by
\bqr
{\cal L} = \phi \Bigl( \Box \phi + g \partial^\beta{}_{\pd}
\partial^\alpha{}_{\pd} \phi \partial_{\beta\pd} \partial_{\alpha\pd}
\phi \Bigr) \ ,
\fqr
and the action they proposed, related by gauge fixing the selfduality
equations differently to contain the mixed derivatives
$\partial_{\alpha\pd} \partial_{\beta\md}$, produce vanishing
tree-level amplitudes in accord with those  of the $\cN=2$ string;
however, they lack Lorentz invariance and require a dimensionful coupling
constant. The quantization is problematic for these reasons. Different
selfdual gravity and selfdual YM actions have been proposed and
quantized which agree with this  previous analysis at tree-level:
Covariant versions for gravity are
${\cal L}= e^{\alpha\dt\gamma}\wedge e^\beta{}_{\dt\gamma}\wedge
d\omega_{\alpha\beta}$ in terms of the vierbein and selfdual Lagrange
multiplier one-forms $e^{\alpha\da}$ and $\omega_{\alpha\beta}$
\cite{Siegel:1993wd}, and a higher-derivative version ${\cal L}=
\rho^{\alpha\beta} \wedge R_{\alpha\beta}$ in terms of the selfdual
curvature and Lagrange multiplier two-forms $R_{\alpha\beta}(e)$ and
$\rho^{\alpha\beta}$ \cite{Gates,Chalmers:2000wd}.  Either
has the light-cone gauge-fixed form containing two fields,
\bqr
{\cal L} = {\tilde\phi} \Bigl( \Box \phi + g \partial^\beta{}_{\pd}
\partial^\alpha{}_{\pd} \phi \partial_{\beta\pd}
\partial_{\alpha\pd} \phi
\Bigr) \ .
\label{twosdgravity}
\fqr

Siegel \cite{Siegel:Idontunderstandthisstupidnotation,Siegel:1993wd}
demonstrated how both Lorentz invariance
and dimensional  analysis could be restored in the target space-time
description by incorporating  space-time supersymmetry, although the
bosonic truncation of this proposal is also  Lorentz invariant and
possesses the dimensional analysis to be conformal.  This action was
quantized in \cite{Chalmers:1996rq}, and the S-matrix shown to  agree
with one-loop MHV amplitudes in gauge theory and gravity at one-loop.
The  fact that vertices in selfdual field theories are independent of
helicity allows  spin to be introduced as an internal symmetry. This is
implemented through superselection sectors analogous to Chan-Paton
factors, as in
\cite{Siegel:Istilldontunderstandthisstupidnotation}.
The helicity independence in the target space-time  is due to
spectral flow in the $\cN=2$ string description; additional states may be
incorporated in the $\cN=2$ string that carry fermionic statistics. The
latter are distinguished  from the naive cohomology analysis by only a
line factor. In fact, a simple doubling of  fields to helicities of both signs
is sufficient to restore Lorentz invariance and dimensional  analysis, but
maximal supersymmetry puts all fields of both signs in the same
multiplet.

\subsection{Loops}

These target space-time actions force all diagrams at more than one loop
to vanish \cite{Chalmers:1996rq}, in agreement with the $\cN=2$ string
theory higher-genus scattering amplitudes: The negative-helicity field
appears linearly in the two-field  selfdual action \rf{twosdgravity}, and
thus it counts loops. It appears as an external field once in each
connected tree graph, never  in one-loop graphs, and no higher-loop
diagrams exist \cite{Chalmers:1996rq}. Furthermore, in any
supersymmetric extension the one-loop field theory graphs also vanish, because
helicity independence of the couplings implies exact cancelation between
fermions and bosons in the loop.

Turning to $\cN=2$ string calculations, cancelations of scattering
amplitudes at all genera have been shown through an $\cN=4$
reformulation \cite{Berkovits:1995vy}  as well as through Ward identities
\cite{Junemann:2000xj}, modulo subtleties  associated with contact term
ambiguities, {\it if} modular invariance is assumed.\footnote{The 
Weyl-Petersson integration measure $d^2\tau/\tau_2^2$ on the
torus is analyzed in many works, for example in section IV.A of 
\cite{D'Hoker:1988ta}.}   The three-point
function does not vanish due to kinematics in $d=2+2$ and  is infra-red
divergent \cite{Bonini:1991bf}.  (The ambiguity in defining the  three-point
function due to regularization \cite{Chalmers:2000wd} necessitates
higher-point genus calculations to identify the quantum target
space-time theory.)   Furthermore, the genus-one four-point function in
the closed $\cN=2$ string  has been examined in detail in
\cite{Chalmers:2000wd} together with a mapping  at $n$-point to this
genus order and an implementation of line factors for  covariance of the
scattering; agreement with the vanishing theorems of
\cite{Berkovits:1995vy} is found, again assuming modular invariance.  This
cancelation is most naturally explained in terms of a four-dimensional
supersymmetric selfdual target space-time theory.  At genus one the
vanishing of the $\cN=2$ string amplitudes was demonstrated directly by
a calculation of the amplitudes in the RNS $\cN=2$ formulation
\cite{Chalmers:2000wd} (including only a single massless scalar degree of
freedom in the spectrum), which demonstrated at the integrand level
that the reason was an additional factor of $\tau_2$ associated
with the ghost system of the world-sheet gauge field
\cite{Marcus:1992wi,Chalmers:2000wd}\footnote{The path integral
quantization at arbitrary genus is analyzed in \cite{Mathur:1988uk}.}.  The
quantum four-point amplitude has been examined at the level of ordering
of limits ($\alpha'$ small and spin structure summation) as well as
contact term interactions on the world-sheet in \cite{Chalmers:2000wd}
and interpretations of the quantum target theory in the $\cN=2$ string
are given, the one-loop amplitudes being identical to the dimensionally
reduced MHV amplitudes.\footnote{An explicit relation between the $\cN=2$
string amplitude at genus one and the ultra-violet portion of the IIB
supergravity non-MHV amplitude in $d=10$ was also found through a dimension
shifting relation.}  In this work we re-examine the quantum scattering and
its consistency in different orders of perturbation theory via the inclusion
of a supersymmetric multiplet of states in the massless spectrum through the
construction of the superselection sectors.  (A previous attempt towards a
supersymmetric extension based on a $Z_2$ twisting is presented in
\cite{Bischoff:1995kb}.)

On the other hand, loop amplitudes can be calculated directly, without any
assumption of modular invariance.  The genus one amplitudes
are conveniently evaluated (since the days of lightcone path integrals,
and even earlier with operator methods) by sewing, and this procedure is
equivalent to the field theory one.  The result is known from field theory
methods, and will be re-derived here by string theory methods:  The
bosonic $\cN=2$ string has {\it nonvanishing} one-loop amplitudes through
this approach, while
the super $\cN=2$ string has vanishing ones.  Since modular invariance
requires vanishing amplitudes at all loops (for $n\geq 4$ $n$-point
amplitudes), this implies that the bosonic $\cN=2$ string has a conformal
anomaly.  In other words, supersymmetry is required to cancel the
anomaly, just as SO(32) is required to cancel anomalies in the open
$\cN=1$ superstring.  (Also, a trivial gauge group choice of $SO(2)$ 
cancels the
interactions in the target space-time theory in the open string case and
thus also the anomaly.)  Thus, in this case world-sheet conformal invariance
requires space-time supersymmetry.  Hidden supersymmetry has
previously  been analyzed in \cite{Witten:1978xv} in the context of
two-dimensional models.  Global anomalies were originally considered
in \cite{Witten:1982fp}.

\subsection{Selfduality in field theory}

Several exact sequences of one-loop gauge theory amplitudes, for
example the MHV ones \cite{Bern:1994qk}
(constructed recursively through analyticity requirements and then
derived for an internal quark in \cite{Mahlon:1994si}) and the
gravity analog \cite{Bern:1998xc} have allowed for explicit
comparisons between the $\cN=2$ string theory quantum amplitudes
and those in field theory. Furthermore, selfduality poses an
interesting structure and reformulation of gauge theory as a
perturbative construction around the selfdual point
\cite{Chalmers:1997sg}; this translates, as opposed to an
expansion in loops, to an expansion in helicity
around the maximal helicity configuration.  (Both can be
formulated as coupling constant expansions.)  Amplitudes in non-selfdual
Yang-Mills theory and gravity simplify as the number of helicities  with
the same sign increases. Supersymmetry identities at the tree-level
enforce the vanishing of the non-supersymmetric MHV  amplitudes to this
order, and the next simplest tree amplitude,  next-to-MHV, was
conjectured and proven in \cite{Parke:1986gb}.
Supersymmetry also forces  the one-loop amplitudes with differing
internal virtual states  to be the same up to a sign.

Selfduality of the field equations implies vanishing of the tree-level
amplitudes through a construction of conserved currents in the case of
gravity \cite{Plebanski}, and through a direct map of the classical
scattering to  vanishing amplitudes in gauge theory \cite{Chalmers:1996rq}:
Bardeen demonstrated a relationship of off-shell gauge fields (at tree
level) between selfdual and non-selfdual theories \cite{Bardeen:1996gk}.
The amplitudes at tree level in selfdual theories all vanish. To one loop
the MHV S-matrices are found to describe the quantum scattering of selfdual
field theories \cite{Chalmers:1996rq, Cangemi:1997rx}.  (The relation of these
different selfdual actions at the quantum level is analyzed
in \cite{Chalmers:1996rq},
and the Lorentz covariant versions may be found through a truncation of
non-selfdual gauge theory to the selfdual limit.)

\subsection{Outline}

In section 2 we analyze the spectrum of this string, after including
internal degrees of freedom for the single massless state, as the
supersymmetric gauged extension of selfdual gravity, and analyze its
corresponding BRST cohomology. We show that there is a supersymmetric
completion of  the $\cN=2$ string through the addition of the
factors labeling spin of external states. In section 3 we analyze duality in
the context of ${\cal N}=2$ string scattering. In section 4 we review 
earlier results for genus-one amplitudes.  In section 5 we perform 
sewing
and unitarity constructions of the genus-one amplitude and compare with
known results, obtaining the conformal anomaly.  The last
section contains concluding remarks.

\section{Spectrum}

The minimal characterization of the spectrum contains a single massless
degree of freedom. However, as is the  case in open string theories,
superselection sectors and the line factors of the external legs may be
incorporated  which represent the internal
symmetries of the string.  In the $\cN=2$  string a graded internal Lorentz
symmetry may be introduced with representatives labeling the
spin states of fields with selfdual couplings.  In a background field
formalism, at quadratic order in the quantum fields, for only
maximum-helicity background fields and lower-helicity quantum fields,
the action has the form of a non-selfdual action in a selfdual background.
It is  known that half-integral spin fields may be ``bosonized" in the
target space-time around such a selfdual field configuration. In this
section we examine the  target space-time supersymmetrization of the
$\cN=2$ string.

\subsection{Bosonization and second-order formulation}

In this subsection we summarize briefly the spin statistics of particles in a
selfdual background.  The action $S = {\rm Tr} \int d^4 x\ {\cal L}$ 
for a minimally coupled fermion $\psi^\alpha$
and its gauge conjugate $\xi_\alpha$ is
\bqr
{\cal L} = \psi^\alpha \nabla_{\alpha\da}
{\bar\xi}^{\da} + c.c. +
    m\left(\psi^\alpha \xi_\alpha + {\bar\psi}^{\da}
{\bar\xi}_{\da} \right) \ ,
\fqr
with $\nabla_{\alpha\db} = i\partial_{\alpha\db} +
A_{\alpha\db}$, the covariant  derivative. Functionally integrating
out the dotted fields gives rise to a  second-order form for the fermionic
couplings \cite{Morgan:1995te},
\bqr
{\cal L} = -{1\over m} \psi^\alpha \left(\nabla^{\alpha\da}
\nabla_{\alpha\da}
    - m^2\right) \xi_\alpha + {1\over m}\psi^\alpha \xi^\beta
F_{\alpha\beta} \ ,
\fqr
where $F_{\alpha\beta}$ is the selfdual projection of the field-strength.
In an anti-selfdual background the fermionic coupling becomes that of a
scalar. In the massless limit (after appropriately scaling the fields with
the mass), the fermionic coupling in such backgrounds is
\bqr
{\cal L} = \psi^\alpha \Box \xi_\alpha \ ,
\label{secondorder}
\fqr
and the index on the field only enters into amplitude calculations through an
external line factor. In this formulation the fermion can be said to 
be ``bosonized", and
there is a direct analog in the $\cN=2$ string. In the supersymmetric
theory, the single spinor index may be represented as an internal
symmetry factor
(i.e., a line factor associated with the Lorentz group representations)
attached to the single creation operator arising in the cohomology analysis.

\subsection{Supersymmetric selfdual gauged supergravity}

The supersymmetric extension of selfdual systems in the context of $N$-extended
selfdual supergravity theory \cite{Siegel:1993wd} is described in the
following.
The target space-time supersymmetric extension of the $\cN=2$ string has the
particle content of this supersymmetric system. This theory is described
by the local $OSp(N\vert 2)$ algebra tensored with the $SL(2)'$ half of the
Lorentz algebra:
\bqr
[\nabla_{A\da}, \nabla_{B\db} \} = {1\over 2} C_{\da\db}
F_{ABCD} M^{CD} \ .
\fqr
The supercoordinate is $z^{M\dt\mu}=(x^{\mu\mu'},\theta^{m\dt\mu})$,
and the
$OSp(N\vert 2)$ superindex $A=(a,\alpha)$ contains the $SL(2)$ index
``$\alpha$" and a vector $SO(N)$ one ``$a$." The covariant derivatives are
defined by
\bqr
\nabla_{A\da}=E_{A\da}{}^{M\dt\mu}\partial_{M\dt\mu} +
    {1\over 2} \Omega_{A\da BC} M^{CB} \ ,
\fqr
where letters from the beginning of the alphabet denote flat local
tangent space indices and those from the middle the coordinates. The
coupling constants are defined as part of the $OSp(N\vert 2)$ metric:
$\eta^{ab}=g \delta^{ab}$, $\eta^{\alpha\beta}= \kappa C^{\alpha\beta}$,
$\eta^{a\beta}=0$. The gravitational coupling constant $\kappa$ has
dimensions of inverse mass.

The light-cone gauge $\nabla_{A{\pd}}=\partial_{A{\pd}}$ solves
the $(\da,\db)
=({\pd},{\pd})$ constraint. The $({\pd},{\md})$ and
$({\md},{\pd})$
constraints may be solved through the introduction of a scalar
superfield $\Phi$ leading
to the covariant derivative $\nabla_{A {\md}}$ given by
\bqr
E_{A{\md}}{}^{M{\dt\mu}}\partial_{M{\dt\mu}}=\partial_{A{\md}} +
    \bigl( \partial_{A{\pd}} \partial_{B{\pd}} \Phi\bigr)
\eta^{CB} \partial_{C{\pd}}
\fqr
\bqr
\Omega_{A{\md}BC} = \partial_{A{\pd}} \partial_{B\pd}
\partial_{C{\pd}} \Phi \ ,
\fqr
and the non-vanishing field strength
\bqr
F_{ABCD}=-\partial_{A{\pd}}\partial_{B{\pd}}\partial_{C{\pd}}
    \partial_{D{\pd}}\Phi \ .
\fqr
The remaining constraint $[\nabla_{A{\md}},\nabla_{B{\md}}\}=0$
gives rise to the field equation
\bqr
\partial_A{}^{\da}\partial_{B\da} \Phi+
\eta^{DC}(\partial_A{}_{\pd}\partial_{C{\pd}}\Phi)
    (\partial_{D{\pd}} \partial_{B{\pd}}\Phi)=0 \ .
\label{fieldeq}
\fqr
As further noted in \cite{Siegel:1993wd}, within the field equation
(\ref{fieldeq})
we may solve for all of the $\theta^{a{\md}}$ dependence
explicitly by examining
the $A=a$ and $B=b$ components: In the superfield $\Phi$ we may take
$\theta^{a{\md}}=0$.

The scalar superfield $\Phi$ expanded in components is then
\bqr
\Phi=\phi+\theta^a \psi_a+\theta^a\theta^b\phi_{ab}+\ldots \ .
\label{superfield}
\fqr
Because only $\theta^{a{\pd}}$ appears in the explicit theta expansion
we will drop  the index $\pd$ in the paper from the supercoordinate and
have it implied. The lowest  component in the theta expansion of
(\ref{superfield}), $\phi$, is the $-2$ helicity  graviton state, which is a
singlet under the gauged $SO(N)$. The components $\phi_{ab}$  are the
$-1$ helicity $SO(N)$ gluons in the adjoint representation
$(\phi_{ab}=-\phi_{ba})$.  The remaining states in the superfield
expansion (\ref{superfield}) are in representations labeled by
anti-symmetric products of $SO(N)$ fundamental indices. There are a
total of $2^{N-1}$ bosonic and $2^{N-1}$ fermionic degrees of freedom
contained in the superfield (\ref{superfield}).

The two-field light-cone superspace action is found from the field equation
(\ref{fieldeq}) by incorporating a Lagrange multiplier field $\tilde\Phi$
conjugate to $\Phi$ and contracting the components $A=\alpha$ and $B=\beta$
with $C^{\alpha\beta}$,
\bqr
S = {\rm Tr} \int d^4x \ d^N\theta~ \Bigl[ {\tilde\Phi} \Box \Phi
    + \eta^{BA} {\tilde \Phi}\bigl( \partial^\alpha{}_{{\pd}}
    \partial_{A{\pd}}\Phi) (\partial_{B{\pd}} \partial_{\alpha
{\pd}}\Phi) \Bigr] \ ,
\label{lightcone}
\fqr
with the superspace measure
$d^N\theta = \prod_{j=1}^N
d\theta^j$ (or an action related by exchanging
$\partial^\alpha{}_{\pd}$ with $\partial^\alpha{}_{\md}$).
For N=8 the two superfields are identified.  The
interactions  in (\ref{lightcone}) have two types of derivative structures
for the types of couplings  from $\eta^{AB}=(\kappa C^{\alpha\beta},
g\delta^{ab})$. All of the interactions  proportional separately to $g$ in
(\ref{lightcone}) are of the same form and are  distinguished only by their
respective group theory factors, and similarly for those  proportional to
$\kappa$. Via spectral flow and the field theory bosonization  described
previously the interactions of the different spin fields may be
incorporated  in the $\cN=2$ supersymmetric string through the addition
of line factors attached to the insertion of local vertex operators.

\subsection{Supersymmetry and spectrum}

We next analyze the superconformal algebra of the $\cN=2$ system and
demonstrate the  supersymmetric spectrum. The generators of the
algebra are spanned by $T$, $G^{\pm}$  and a $U(1)$ current $J$ that may
be constructed by the usual Noether method on the gauge-fixed
$\cN=2$ world-sheet action, including the ghost systems $(b,c)$,
$(\beta^\pm,\gamma^\pm)$, and $(b',c')$. The bosonized form of the
ghost systems are
\bqr
c= e^\sigma \qquad b=e^{-\sigma}
\fqr
\bqr
c'=e^{\sigma'} \qquad b'=e^{-\sigma'}
\fqr
for the reparameterizations and $U(1)$ gauge invariance, and
\bqr
\beta^\pm(z) = \partial_z \chi^\pm e^{-\phi^\mp} \qquad
\gamma^\pm = \eta^\pm e^{\phi^\pm}
\fqr
for the doubled holomorphic supersymmetries. Further bosonization of
the fermionic $\chi^\pm$ and $\eta^\pm$ are possible, but unnecessary
for this work.

Also, the system possesses a gauged $U_A(1)$ symmetry.  Its holomorphic
representative is generated by the operator
\bqr
\Sigma (\theta) = \exp\left[ 2\pi\theta \oint {dz\over 2\pi i} \ln(z) 
J(z)\right]
= e^{2\pi i\theta {\hat\phi}(z)} \ ,
\label{sfo}
\fqr
where ${\hat\phi}(z)= \phi^+-\phi^-+\psi^+-\psi^- + cb'$, the
bosonized form of the complete $U(1)$ current $J=\pa {\hat\phi}$. The
operator in \rf{sfo} is essentially a continuous  modification of a spin field
in the usual $\cN=1$ string. It possesses  an inverse, $\Sigma(-\theta)$.
The insertion in a string amplitude of
$\Sigma(\theta) \Sigma(-\theta)=1$, by contour deformation, leads to a
branch cut of monodromy $e^{2\pi i\theta}$ between two chosen points
$z_1$ and $z_2$ on the Riemann surface.\footnote{Such a branch may be
resolved by a $1/\theta$ covering of the Riemann surface, which is itself
a higher genus surface found via a multiple covering of the original
punctured surface.  Thus spectral flow also
is compatible with the triviality of the
S-matrix in the $\cN=2$  theory.} The spectral flow operator in \rf{sfo} at
$\theta=1/2$  connects the Neveu-Schwarz vacuum to the Ramond
vacuum and generates target  space-time fermionic statistics when
placed at the point of emission of a  bosonic state.

\begin{figure}
\begin{center}
\epsfig{file=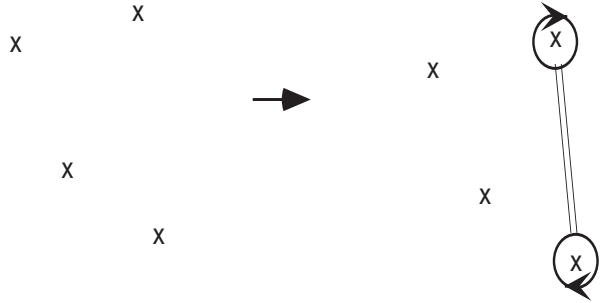,height=5cm,width=8cm}
\end{center}
\caption{Contour deformation of the insertion of
$\Sigma(\theta)\Sigma(-\theta)$.}
\end{figure}

Furthermore, an automorphism of the superconformal
algebra\footnote{We denote currents in the critical case with a hat, e.g.
${\hat G}^\pm$.} on the holomorphic currents is generated through
\bqr
\Sigma(\theta) T(z) \Sigma(-\theta) = T(z) + {\theta\over z} J(z) +
{c\over 6} {\theta^2 \over z^2}
\fqr
\bqr
\Sigma(\theta) G^\pm \Sigma(-\theta) = z^{\mp \theta} G^\pm(z)
\fqr
and
\bqr
\Sigma(\theta) J(z) \Sigma(-\theta) = J(z) + {c\over 3} {\theta\over z} \ .
\label{automorphism}
\fqr
At criticality $c=c_m+c_{gh}=6-6=0$ and we see the $U(1)$ current to be
invariant and that the revolution by $\theta$ twists the supercurrents
$G^\pm$ with the same monodromy due to a $\theta$ moded spin field.

The holomorphic BRST current on the super world-sheet (and
anti-holomorphic relative) is given by
\bqr
J_{BRST} &=& c T + \gamma^+ G^- + \gamma^- G^+ + c' J + c \pa c b + c \pa
c' b' - 4\gamma^+\gamma^- b
\cr &&
+ 2\pa \gamma^- \gamma^+ b' - 2\pa\gamma^+
    \gamma^- b'
+ \sfrac34 \pa c\left( \gamma^+ \beta^- + \gamma^-\beta^+\right)
\cr &&
- \sfrac34 c \left( \pa\gamma^+ \beta^- +\pa\gamma^- \beta^+\right)
+ \sfrac14 c \left( \gamma^+\pa\beta^- +\gamma^-\pa\beta^+\right)
\cr &&
+ c' \left(\gamma^+\beta^--\pa \gamma^-\beta^+\right) \ ,
\fqr
together with ghost number current
\bqr
J_{gh} = -bc -b'c'+\eta^+\xi^- +\eta^-\xi^+ \ ,
\fqr
and a conserved BRST charge of
\bqr
Q= \oint {dz\over 2\pi i} J_{BRST} \ .
\fqr
We shall not list all of the various transformations due to BRST of the
currents but point out the relations
\bqr
[Q,\Sigma(\theta)] = 0 \qquad \pa_z \Sigma(\theta) = -\theta \left\{
    Q,b'(z) \Sigma(\theta) \right\} \ ,
\fqr
which illustrates that $\Sigma(\theta)$ is BRST-closed and that
$\pa_z \Sigma(z)$
is exact.

Together from the ghost system and the BRST charge we define the picture
changing operators
\bqr
PCO^\pm = \left\{ Q,\xi^\pm \right\}
\fqr
required at higher-genus for modular invariance \cite{Friedan:1986ge},
which are also represented on the world-sheet by insertions of the
rebosonized operators
\bqr
PCO^\pm(z) = \delta(\beta^\pm(z)) {\hat G}^\pm(z) \ .
\fqr
The spectral flow operator $\Sigma(\theta)$ commutes with $PCO^\pm$
and shifts the picture number via
\bqr
(\pi^+,\pi^-) \rightarrow (\pi^+ +\theta,\pi^- - \theta) \ .
\fqr
We next re-examine the states of the $\cN=2$ string in the ``NS-sector" and
the ``R-sector," related to each other by spectrol flow operations of
$\Sigma({\sfrac12})$.

In the canonical formulation we denote ground-states of picture number
(of two supercurrents) with
\bqr
\vert \pi^+, \pi^-; k \rangle \ ,
\fqr
which are related to each other through actions of $PCO^\pm$, shifting
$\pi^\pm\rightarrow \pi^\pm+1$, respectively. The single state, without internal
line factors attached, is found in \cite{Junemann:1999rx} as
\bqr
c_1\vert -1,-1;k\rangle
\fqr
with the on-shell condition $k^2=0$, which is the unique state at ghost
number equal to one. The (holomorphic) vertex operator constructing this 
state is
\bqr
V_{-1,-1}^{(1)} = c e^{-\phi^+-\phi^-} e^{ik\cdot x} \ ,
\fqr
and may be transformed into a graviton vertex operator of polarization
$++$  found in the following after normalizing the line factor
\cite{Chalmers:2000wd}.

The vertex operators produce from the (first-quantized) vacuum state the
asymptotic string states in the scattering amplitude under consideration.
They correspond to the physical states of the $\cN=2$ closed string
and carry their quantum numbers. Being representatives of the (semi-chiral 
and semi-relative) BRST cohomology, they are unique up to BRST-trivial 
terms and normalization due to internal symmetry factors. The physical 
subspace of the $\cN=2$ string Fock space through the covariant quantization
scheme contains the ground state $|k\rangle$, a scalar on the massless level,
i.e., for center-of-mass momentum~$k^{\pm a}$ with $k\cdot k=0$. This is not
including internal symmetry factors and states related to the ground state
through spectral flow. The canonical massless scalar field is denoted by
\bqr
\Phi(x)\ =\ \int\!\!d^4k\;e^{-i k\cdot x}\;\tilde{\Phi}(k) \ ,
\fqr
which allows for a supersymmetric completion.  We may relate the higher
states in the supermultiplet to the ground state through spectral flow
and the isospin labeling of the Lorentz group through \rf{superfield}.

The field representative of the vertex operator for a selfdual graviton,
with polarization $\eps_{\mu\nu}^{++}$, is in the $(0,0)$ pictures
\bqr
V(k)= \int d^2z \sqrt{g} ~
\epsilon^{++}_{\mu\nu} \cdot \left( \pa x^\mu + i\psi^\mu k\cdot \psi\right)
\left( {\bar\pa} x^\nu + i{\bar\psi}^\nu k\cdot {\bar\psi} 
 \right) 
 e^{ik\cdot x} \ ,
\label{vertex}
\fqr
and that for a holomorphic space-time fermion in the $(\sfrac12)$
pictures,
\bqr
V_F(k) =\int dz\ L_\mu \left( 
 \partial x^\mu + {i\over 4} k\cdot \psi \psi^\mu \right) e^{ik\cdot x} 
+ {1\over 2} u^\alpha e^{3\phi / 2} b \eta S_\alpha e^{ik\cdot x} 
\fqr
with $L$ a spin operator that connects the NS to the R vacua (the latter 
term does not contribute to the amplitudes at genus one but is necessary 
for BRST invariance).  In the $\cN=1$ superstring this line factor is 
composed, after the bosonization, as
\bqr
L_\mu = u_\alpha \Gamma_\mu^{\alpha\beta} \Sigma_{-1/2} S_\beta \ ,
\fqr
with spin fields
\bqr
\Sigma_{\pm 1/2} = e^{\pm i \sigma(z)/2} \qquad
\beta = \partial_z \chi ~ e^{i\sigma} \qquad \gamma=\eta e^{-i\sigma}
\fqr
   (note also that $\left\{ Q_\alpha, Q_\beta\right\} =
\Gamma_\mu^{\alpha\beta} P^\mu$) and the spin operator from the
internal bosonized world-sheet fermions,
\bqr
S_\alpha = e^{\pm i\phi_1/2 \pm i\phi_2/2 +\ldots} \ .
\label{Soperator}
\fqr
Furthermore, on the world-sheet supersymmetry generators are found by
\bqr
Q_\alpha = \oint {dz\over 2\pi i} V_\alpha = \oint {dz\over 2\pi i}
    \Sigma_{-1/2} S_\alpha \ ,
\label{super}
\fqr
at zero-momentum. The index $\alpha$ denotes the different $\pm$ signs
in the exponential ($2^{d/2}$ for $d$ even) and the world-sheet fermions
bosonized  into $\psi_\pm^a = e^{\pm i\phi^a}$ after complexifying pairs
of them into $\psi_\pm^a = \psi^a\pm i \psi^{d/2+a}$. Currents are $J_a =
-i\partial_z \phi_a$.\footnote{In the $(-1/2)$ picture the fermionic vertex
operator has  the form $V_F = \Sigma_{1/2}u^\alpha S_\alpha e^{ik\cdot x}$.}
The gauge-fixed $\cN=1$ string has a global $\cN=2$ supersymmetry that
allows a direct connection  with the vertex operator construction in the
$\cN=2$ superstring.

There is an additional $\beta\gamma$ ghost system in the $\cN=2$
string because  there are twice as many local supersymmetries (denoted
by $\beta^\pm\gamma^\pm$).  The fact that fermions can be bosonized
in the selfdual background means that  the square-root monodromies
inserted into the string world-sheet transfer  into normalizations after
doing the same for the vertex operator in the $\cN=1$  string. This can be
made manifest in the $\cN=2$ supersymmetric string due to  spectral flow,
i.e., the presence of an additional gauge current on the world-sheet  at
criticality.

In the $\cN=2$ superstring the vertex operator for the emission of a
space-time fermion may be found by taking every component of the
previous bosonized
form of the world-sheet fermionic contributions and gauging the spin
operator  via the $U(1)$ current. The four-component spin operator in
$d=2+2$ target space-time is
\bqr
S_+ = e^{i\phi_1/2 +i\phi_2/2} \qquad S_- = e^{i\phi_1/2-i\phi_2/2}
\fqr
\bqr
S_{\pd} = e^{-i\phi_1/2-i\phi_2/2} \qquad S_{\md} =
e^{-i\phi_1/2+i\phi_2/2} \ .
\fqr
Gauging the expanded component form of the above fermionic vertex
operator with a $U(1)$ current allows a twist of the vertex operator
\bqr
    \Sigma({1/2}) u_\alpha(k) \Gamma_\mu^{\alpha\beta}
  \Sigma_{-1/2} S_\beta
    = \epsilon^+_\mu (k) \ ,
\label{momentumtwist}
\fqr
which amounts to a momentum dependent twist of the single vertex operator
into that of a space-time boson. The $\cN=2$ string carries this in a
straightforward
way at the level of the cohomology through spectral flow and allows internal
quantum numbers to be assigned to states into a priori independent sectors
related by the spectral flow. The translation introduces an
additional redundancy
into the components of the $\eps_\mu^+=\eps_{\alpha\da}^+$ and allows
this twisted line factor to be reduced further to a two-component spinor
$v_\alpha$. The physical polarization in \rf{momentumtwist} has the explicit
bi-spinor form
\bqr
\eps_{\alpha\da}^+(k;q) = {q_\alpha k_{\da}\over
q^\beta k_\beta} \ ,
\fqr
with $q_\alpha$ arbitrary (the reference momenta \cite{ChineseMagic}) and
$\eps^\pm\cdot \eps^\mp=-1$ and $\eps^\pm\cdot\eps^\pm=0$. Different
choices of $q_i$ generate axial forms,
\bqr
\eps_{\alpha\da}^+(k;q_1) = \eps_{\alpha\da}^+(k;q_2) +
f(k;q_1,q_2) k_{\alpha\da} \ ,
\fqr
and are the same on shell due to transversality for on-shell amplitudes.  An
analogous spinor reference momentum is incorporated into the line factor
for  the spinor in \rf{secondorder} \cite{Morgan:1995te}.

The two vertex operators are similar, except that in the $\cN=1$ string
the factor $L^\mu$ introduces a branch cut on the world-sheet and
$\eps^{++}$  a Lorentz covariant line factor (introduced into the $\cN=2$
string following  spinor helicity techniques in \cite{Chalmers:2000wd}). In
a selfdual background  the two are the same up to a line factor, as
required by known field theory  results of bosonization.

A spectral flow operation in a scattering process allows the identification
\bqr
\langle \ldots V_{F}(k_1) V_{F}(k_2) \ldots \rangle
&=&
\langle \ldots V_{F}(k_1) \Sigma(-\theta) \Sigma(\theta) V_{F}(k_2)
\ldots \rangle \cr
&=&
\langle \ldots V_{b}(k_1) V_{b}(k_2) \ldots \rangle \vert_{\theta=1/2} \ ,
\fqr
where in the latter a contour deformation of the spin operator is taken
around the two vertex operators to convert them to bosonized statistics
(see Figure 1).  A similar procedure may be performed for all correlators
involving an even number of fermionic vertex operators. This is the world-sheet
derivation of the bosonization of the fermions in a selfdual background.

In the supersymmetric target space-time theory the quantum amplitudes
are zero to all orders in the field theory limit due to a supersymmetric
identity of MHV amplitudes (originally found in \cite{superidentity}) in
both the closed supersymmetric  $\cN=2$ string as well as the open
$\cN=2$ supersymmetric string. The $\cN=2$ superstring,  found by
multiplying the Fock space with internal factors labeling spin, agrees to
all  orders with the vanishing theorems of the $\cN=2$ string that
assumed modular invariance.

\section{Duality implies vanishing}

In the early days of string theory many properties of amplitudes were
shown to follow from (Dolen-Horn-Schmid) duality \cite{Dolen:1968jr} (via
``stretching the worldsheet") and without identification as any particular
string theory. For example, duality alone shows that all loop diagrams
can be expressed as insertions of tadpoles into trees \cite{Lovelace}.  We
analyze this duality within the gauge theory MHV amplitudes in the
following.

For a theory with a finite number of particles of generic mass, duality is
a much more stringent constraint: Since a finite sum of poles in one channel
cannot equal such a sum in another channel, the tree amplitudes must all
vanish (except for the three-point function, which does not have any
channels).
This analysis can be extended to one-loop diagrams: For example,
a planar one-loop open-string graph is related by duality to such a tree
graph with the insertion of a closed-string tadpole. The MHV one-loop YM
amplitude \cite{Bern:1994qk,Mahlon:1994si}, which coincides with the one-loop
selfdual S-matrix, has the
form
\bqr
\cA_{n;1}^{[1]}(k_i)\ =\
-{i\over 48\pi^2} \sum_{1\leq i<j<k<l\leq n} {\langle ij\rangle
[ jk] \langle kl\rangle [ li] \over
\langle 12\rangle \langle 23\rangle \cdots \langle n1\rangle}
\label{allplusgauge}
\fqr
in a color-ordered form.  Relations between this amplitude in $d$ dimensions
and a supersymmetric amplitude in $d+4$ dimensions relates scattering between
the $\cN=1$ and $\cN=2$ strings \cite{Chalmers:2000wd}.  The inner products are
written in terms of twistor variables $\langle ij\rangle =
k_i^\alpha k_{j\alpha}$
and $[ij] = k_i^{\da} k_{j\da}$, with $s_{ij} = (k_i+k_j)^2
= \langle ij\rangle [ji]$. In $d=2+2$ the amplitude becomes real and
the inner products $\langle ij\rangle$ and $[ij]$ are not complex
conjugates. A similar result holds for the MHV gravitational amplitudes
\cite{Bern:1998xc}. The form in \rf{allplusgauge} is manifestly crossing
symmetric ($i\leftrightarrow j$) after including the color factor
\bqr
C_\sigma = {\rm Tr} ~ T^{\sigma_1} T^{\sigma_2}\cdots T^{\sigma_n}
\fqr
with group generators $T_i$ and summing over all permutations, as all the
outgoing helicities are the same.

The analytic structure (in momentum space) of a closed tree diagram with
zero-momentum inserted states, and an open loop diagram containing a
finite number of particle types, in general do not agree. We next analyze
the insertion of a single operator in a field theory tree diagram. The
(color ordered) gauge field corresponding to a tree diagram with  $m$
out-going helicities the same (labeled from legs $1$ to $m$) and  one leg
unamputated and off-shell is
\bqr
A_m^{\alpha\db}(q;k_i^+) =
{k^\alpha q^{\gamma\db}k_\gamma
\over \langle k1\rangle \langle 12\rangle \cdots
\langle mk\rangle} \ ,
\fqr
with $k$ the reference momentum \cite{ChineseMagic} chosen in the
off-shell extension of the leg with momentum $q$ ($=-\sum_{j=1}^m k_j$)
\cite{Parke:1986gb}.
The insertion of a zero-momentum state interacting with a vertex associated
with $j$ lines requires products of up to $j$ fields. The overall constant
$c_m$ is suppressed in the subsequent analysis.  These composite operator
insertions model local anomaly mechanisms; as an example we list
\bqr
S = {\rm Tr}~ \int d^4x ~ \phi F {\tilde F}  \ .
\fqr

We denote the product of two fields as
\bqr
A^{\alpha\db}_m A^{\delta\dt\rho}_{n-m} =  {1\over
\langle 12\rangle \langle 23\rangle \cdots \langle n1\rangle}
    H^{\alpha\db,\delta\dt\rho} \ ,
\fqr
and consider the general contraction with two local derivatives with momentum
$q$,
\bqr
\cA= q^{\alpha_1\db_1} q^{\alpha_2\db_2}
H^{\alpha_3\db_3,\alpha_4\db_4}
    T_{\alpha_1\db_1,\alpha_2\db_2,\alpha_3\db_3,\alpha_4\db_4} \ ,
\fqr
with $T$ a general tensor. As the momentum $p$ flowing into the product of two
fields is taken to zero, the field is evaluated with momentum $q$ and $-q$.
The unamputated and off-shell field is conserved via $q^{\alpha\db}
A_{\alpha\db}=0$,
and the most general tensor contractions are obtained from the tensors,
\bqr
T_{(1)}^{\alpha_1\db_1,\alpha_2\db_2,\alpha_3\db_3,\alpha_4\db_4} =
\epsilon^{\alpha_1\alpha_3} \epsilon^{\alpha_2\alpha_4} \epsilon^{\db_1\db_3}
\epsilon^{\db_2\db_4}
\fqr
\bqr
T_{(2)}^{\alpha_1\db_1,\alpha_2\db_2,\alpha_3\db_3,\alpha_4\db_4} =
\epsilon^{\alpha_1\alpha_4} \epsilon^{\alpha_2\alpha_3} \epsilon^{\db_1\db_4}
\epsilon^{\db_2\db_3} \ ,
\fqr
\bqr
T_{(3)}^{\alpha_1\db_1,\alpha_2\db_2,\alpha_3\db_3,\alpha_4\db_4} =
\epsilon^{\alpha_1\alpha_3} \epsilon^{\alpha_2\alpha_4} \epsilon^{\db_2\db_3}
\epsilon^{\db_1\db_4} \ ,
\fqr
\bqr
T_{(4)}^{\alpha_1\db_1,\alpha_2\db_2,\alpha_3\db_3,\alpha_4\db_4} =
\epsilon^{\alpha_1\alpha_4} \epsilon^{\alpha_2\alpha_3} \epsilon^{\db_1\db_3}
    \epsilon^{\db_2\db_4} \ ,
\fqr
together with those that trivially contribute zero due to momentum
conservation,
such as $\epsilon^{\alpha_1\alpha_2} \epsilon_{\alpha_3\alpha_4}
\epsilon^{\db_1\db_3}
\epsilon^{\db_2\db_4}$ and permutations. The contractions with the
various tensors
after summing over all possible $m$ and $n-m$ point fields generates,
\bqr
\cA_1 = \sum_{i=1}^m \sum_{j=m+1}^n \sum_{a=1}^m \sum_{b=m+1}^n ~
{\langle ka\rangle \langle kb\rangle [ia] [jb] \langle ik\rangle
\langle jk\rangle
\langle m,m+1\rangle \langle n1\rangle \over
\langle k1\rangle \langle mk\rangle \langle k,m+1\rangle \langle nk\rangle}
\label{contractone}
\fqr
\bqr
\cA_2 = \sum_{i=1}^m \sum_{j=m+1}^n \sum_{a=1}^m \sum_{b=m+1}^n ~
{\langle ka\rangle \langle kb\rangle [ib] [ja] \langle ik\rangle
\langle jk\rangle
\langle m,m+1\rangle \langle n1\rangle \over \langle k1\rangle
\langle mk\rangle
\langle k,m+1\rangle \langle nk\rangle} ,
\label{contracttwo}
\fqr
\bqr
\cA_3 = \sum_{i=1}^m \sum_{j=m+1}^n \sum_{a=1}^m \sum_{b=m+1}^n ~
{\langle ka\rangle \langle kb\rangle [ib] [ja] \langle ij\rangle
\langle jk\rangle
\langle m,m+1\rangle \langle n1\rangle \over \langle k1\rangle
\langle mk\rangle
\langle k,m+1\rangle \langle nk\rangle} \ ,
\fqr
and
\bqr
\cA_4 = 0
\fqr
In deriving \rf{contractone} and \rf{contracttwo} the null vector $q$ has
been written as $q^{\alpha\db}= \mathop\sum_{j=1}^m k_j^\alpha
k_j^{\db}$ together with momentum conservation. The first sum in
\rf{contractone} is separately odd under both $i\leftrightarrow a$ and
$j\leftrightarrow b$, and the second is odd under $i\leftrightarrow b$ and
$j\leftrightarrow a$. Both
$\cA_1$ and $\cA_2$ are equal to zero. The series in $\cA_3$ is odd under
$i\leftrightarrow b$ and $j\leftrightarrow a$ separately (and even under
simultaneous
$(i,j)\leftrightarrow (a,b)$).

We next consider the zero-momentum insertion associated with a
three-point vertex.
Summing over all possible products of gauge fields associated with
a general tensor
gives
$$
\cA^{\alpha_1\db_1,\alpha_2\db_2,\alpha_3\db_3} = \sum_{i=1}^m
\sum_{j=m+1}^p \sum_{a=p+1}^n
    k^{\alpha_1} k_i^{\db_1} k^{\alpha_2} k_j^{\db_2} k^{\alpha_3} k_a^{\db_3}
$$
\bq
\times
{\langle ik\rangle \langle jk\rangle \langle ak\rangle \over \langle k1\rangle
\langle m,m+1\rangle \langle p,p+1\rangle \langle n1\rangle \langle mk\rangle
\langle k,m+1\rangle \langle pk\rangle \langle k,p+1\rangle \langle
qk\rangle} \ .
\fq
Any contraction of the indices $\alpha_j$ associated with $k_\alpha$ or a
$\epsilon^{\alpha\beta}$ generates $\langle kk\rangle$ and is identically
equal to zero.

The last possibility contains a four-point vertex,
$$
\cA^{\alpha_1\db_1,\alpha_2\db_2,\alpha_3\db_3,\alpha_4\db_4}
= \sum_{i=1}^m
\sum_{j=m+1}^p \sum_{a=p+1}^q \sum_{b=q+1}^n
k^{\alpha_1} k_i^{\db_1} k^{\alpha_2} k_j^{\db_2} k^{\alpha_3} k_a^{\db_3}
k^{\alpha_4} k_b^{\db_4}
$$
\bq
\times
{\langle ik\rangle \langle jk\rangle \langle ak\rangle \langle bk\rangle
\langle m,m+1\rangle \langle p,p+1\rangle \langle q,q+1\rangle
\langle n1\rangle
\over \langle k1\rangle \langle mk\rangle \langle k,m+1\rangle
\langle pk\rangle
\langle k,p+1\rangle \langle qk\rangle \langle k,q+1\rangle \langle
nk\rangle} \ ,
\fq
the form of which also generates zero for any tensor contraction.

The above analysis shows that the non-supersymmetric gauge theory MHV
amplitudes
cannot be generated by a single zero-momentum operator insertion into
a tree amplitude.  We conclude that the string duality between open and
closed (in the $\cN=2$ system)  requires that the amplitudes vanish.
Selfdual {\it super} Yang-Mills and selfdual {\it  super}gravity are
examples of such theories, but the corresponding nonsupersymmetric
theories are not.

\section{Modular invariance vs.\ field theory}

The evaluation of scattering amplitudes for the string, in either the 
path integral or operator approach, involves
sums over the inequivalent geometries associated with the worldsheet
topology.  There are two steps to this procedure, finding (1) the 
Green functions on that space and (2) the corresponding measure.
The second step is the hardest; in earlier evaluations of one-loop 
quantities in $\cN=2$ string theory, it was {\it assumed} that the 
measure was modular invariant, to enforce conformal
invariance in the critical
dimension of four.  Calculations
with this integration measure led directly to vanishing genus-one diagrams.
This result directly contradicted explicit evaluations of the 
corresponding graphs by field theoretic methods.
In this section we will review these results; a more rigorous 
analysis in the following section, based on sewing or unitarity 
constructions, will reveal  an anomaly in conformal invariance in the 
non-supersymmetric string.

We first examine the consistency of the zero-point function in the
supersymmetric context. At genus one it is given by
\bqr
Z= \int {d^2\tau\over \tau_2^2} = {\pi\over 3}
\fqr
and corresponds in the target space-time field theory (on a $2d$-real
dimensional K\"ahler manifold of signature $(4,0)$ or $(2,2)$~) to
the zero-point function \cite{Ooguri:1991fp} derived from
\bqr
\int_{\cal M} d^dx~ \sqrt{g} = {1\over \left({d/2}\right)!} \int_{\cal M}
    ~ \omega\wedge\omega\ldots \wedge\omega \ ,
\label{cosmological}
\fqr
with a product of ${d/2}$ factors of the K\"ahler form $\omega$. The
integrand in \rf{cosmological} is locally a total derivative, as $\omega=
d{\bar d} K$ (e.g. $d\omega={\bar d}\omega=0$), and the integrand is totally
antisymmetric,
\bqr
\int_{\cal M} d^dx~ \sqrt{g} \sim \int_{\partial M} {\bar d} K\wedge\omega
\wedge\omega\ldots = \int_{{\bar\partial}M} dK\wedge\omega\wedge\omega\ldots
\fqr
The fact that the cosmological term in \rf{cosmological} is a total derivative
(locally) means that it does not contribute to the field equations. The absence
is in agreement with the vanishing of the cosmological term in the selfdual
gravity theory and in the supersymmetric extension.

The four-point one-loop MHV gravitational amplitude in
four-dimensions (in $d=3+1$ dimensions) has the form
\bqr
\cA_{4}^{[2]}(k_i)\ =\
-i \Bigl({\kappa\over 2}\Bigr)^4 {1\over 120\,(4\pi)^2}\, \Bigl(
{s_{12} \, s_{23} \over
\langle12\rangle \langle23\rangle \langle34\rangle \langle41\rangle }
\Bigr)^2 \, (s_{12}^2 + s_{23}^2 + s_{13}^2) \ ,
\label{fourpointgrav}
\fqr
derived in \cite{Grisaru:1980re}. Its $n$-point form is presented in
\cite{Bern:1998xc} and generates the S-matrix for the selfdual
gravitational field theory \cite{Chalmers:1996rq}. The supersymmetric
completion (integrating out a virtual supersymmetric multiplet) is
identically zero, and the amplitude satisfies a relation
$\cA^{[2]}=\cA^{[0]}  =-\cA^{[1/2]}$ for an internal graviton, complex
scalar and Weyl fermion  respectively. The $d$-dimensional form, found by
analytically continuing the internal momenta or by inserting factors of
$\tau_2^{-d/2+2}$  into the first quantized integral form, is equal to zero
in $d=2$ \cite{Chalmers:2000wd}.

The all-genus amplitudes have been shown to be equal to
zero through calculations in the $\cN=4$ topological reformulation
of the $\cN=2$ string (modulo contact term ambiguities analyzed in
\cite{Chalmers:2000wd}), based on the
assumption of modular invariance. This has been  verified directly at
genus one in the RNS $\cN=2$ formulation, taking into  account both
contact terms as well as the different ordering of limits in the  zero-slope
limit \cite{Chalmers:2000wd} (integrating spin structure first or evaluating 
$\alpha'\rightarrow 0$ first), generating the continued form of  the result
in \rf{fourpointgrav} to $d=2$.  In the latter calculation it  was shown that
an additional factor of $\tau_2$ in the closed string  calculation arising
from the $b'c'$ ghost system associated with the  world-sheet gauge
field maps the integral representations of the amplitude into two internal
space-time dimensions, and the result for the  amplitude equals zero.
However, this additional factor in the integration  measure does not
follow directly from sewing trees or from unitarity  considerations (in the
Wick-rotated sense to $d=3+1$ dimensions); these latter techniques would
generate the non-vanishing gravitational MHV amplitudes
\rf{fourpointgrav} at $n$-point.

The chiral $\cN=2$ matter multiplets $X=(x,\psi)$ and ghost systems at
genus one contributes the following determinant factors to the 
evaluation of the
amplitude:
\bqr
Z_d\ab(\tau,\bar\tau)\ =\
Z_x(\tau,\bar\tau)\, Z_{\psi}\ab(\tau,\bar\tau)\, Z_{bc}(\tau,\bar\tau)\,
Z_{\beta\gamma}\ab(\tau,\bar\tau)\, Z_{b'c'}(\tau,\bar\tau)
\ ,
\fqr
with $d=4$ in the critical string and where the respective factors are
\bqr
Z_x(\tau,\bar\tau)\ =\ \tau_2^{-d/2} \vert\eta(\tau)\vert^{-2d} \qquad\qquad
Z_\psi\ab(\tau,\bar\tau)\ =\
\vert \vartheta\ab(0,\tau)\vert^d\, \vert \eta(\tau) \vert^{-d} \ ,
\label{matterfields}
\fqr
\bqr
Z_{bc} (\tau,\bar\tau)\ =\ \tau_2\,\vert \eta(\tau)\vert^4 \qquad\qquad
Z_{\beta\gamma}\ab(\tau,\bar\tau)\ =\
\vert \vartheta\ab(0,\tau)\vert^{-4}\, \vert \eta(\tau) \vert^4
\fqr
The ghost determinant associated with the local $U(1)$ symmetry is,
\bqr
Z_{b'c'}(\tau,\bar\tau)\ =\ \tau_2\,\vert \eta(\tau)\vert^4  \ .
\label{ghostdet}
\fqr
The Dedekind eta and theta functions with continuous characteristic
$\ab$ comprise the determinants,
\bqr
\eta(\tau)\ =\ q^{1/24} \prod_{n\neq 0} (1-q^n)  ~\, \qquad
\vartheta\ab (z,\tau)\ =\
\sum_{n\in Z} e^{\pi i \tau (n+\alpha)^2
  + 2\pi i(n+\alpha)(z+\beta)} \ ,
\fqr
where $q=e^{2\pi i\tau}$ and $\tau$ denoting the modular parameter of the
torus.  The product of these factors generates,
\bqr
Z_d\ab(\tau,\bar\tau)\ =\ \tau_2^{-{(d-4)\over 2}}\,
\vert \vartheta\ab(0,\tau)\vert^{d-4}\,
\vert \eta(\tau)\vert^{-3(d-4)} \ ,
\label{fulldetset}
\fqr
and equals unity in four real dimensions.  The moduli associated with
the path integral quantization of the $\cN=2$ string are : 1) the
parameter $\tau$ labeling the inequivalent tori, and 2) the parameters
$\ab$ labeling the continuous spin structures (or the $U(1)$ gauge bundle
of the torus).  The modular invariant integration measure is (up to 
an arbitrary constant)
\bqr
{d^2\tau\over \tau_2^2}  \qquad\quad  {du d{\bar u}\over \tau_2}
  =d\alpha d\beta \ ,
\label{modmeasures}
\fqr
where $u=(1/2-\alpha)+(1/2-\beta)\tau$.  (The volume integral is $\int du
d{\bar u} =\tau_2$.)  However, this is not the measure found from 
path integral (or operator) quantization, as we show in the following 
section.  

\section{Anomaly in $\cN$ = 2 string}

A {\it rigorous} alternative to invariance arguments is
``sewing", which {\it unambiguously} determines loop amplitudes
inductively in loop order.  This property is an automatic consequence of
the path integral because of the locality of the world-sheet action:  For
example, the genus-one worldsheet can be ``cut" into trees, and
integration over variables at the cut are performed after integration over
those inside the tree.  (Similar remarks apply in the operator formalism.)
This is equivalent to the Feynman tree theorem \cite{Feynman}, and is the
original method for evaluating one-loop graphs in string theory (see, e.g.,
\cite{Mandelstam:1968hz,Kikkawa:1969qy}).  The basic idea is that the
one-loop graph is given by the trace of the string propagator (with
additional external states).

For example, this method was applied to the Type I superstring to show
the existence of anomalies (and divergences) for groups other than
SO(32).  Although the integration measure is determined directly from
the Teichm\"uller variations,
conformal invariance is not enough to fix the normalization of these
one-loop diagrams:  In fact, it gives an inconsistent result, since the
theories for groups other than SO(32) are anomalous; assuming no
anomaly would imply different relative normalizations to enforce
cancelation.  In particular, ignoring the possibility of Chan-Paton factors
would give the U(1) or SO(2) (if symmetrized) superstring, which does
violate conformal invariance at one loop.  Of course, conformal invariance
does not fix overall normalization of loop diagrams in any string theory,
since any constant overall factor is invariant; rather unitarity determines
the normalization factor.

The simplest case to consider is the open string planar loop, since the
moduli space of the resulting integral is simpler (at least for the part
coming from the worldsheet metric).  Specifically, the propagator for an
open string in external fields is simply
$$
\Delta = {1\over L_0 +V} = {1\over L_0} -{1\over L_0}V{1\over L_0} +...
$$
where $L_0$ is the free kinetic operator and $V$ is the vertex
operator(s).  The loop amplitude, for some fixed number of external lines,
is then given by sewing a term in this sum, which is represented in this
operator language as a trace:
$$
{\cal A} = Tr \left( {1\over L_0} V(1) {1\over L_0} V(2) ... \right)
$$

At this starting point there are no moduli whatsoever, since
the tree-level path integral has already been performed.  In general,
moduli appear only as integration variables, and do not specify external
states.  However, it is convenient to reintroduce one modulus for
purpose of evaluating the sewing:  Exponentiating all the free
propagators $1/L_0$ with the usual Schwinger parameters, this modulus
appears as their sum.

We will not review all the details here, just those that differ from the
$\cN=0$ and $\cN=1$ strings.  (See, e.g., \cite{GSW} for a discussion for the
bosonic open string.)  For simplicity, we can consider Neveu-Schwarz or Ramond
boundary conditions for the worldsheet spinors:  This allows direct
comparison to the $\cN=1$ case.  (By spectral flow, also known in the 
maximally helicity violating sector as spacetime supersymmetry, we 
know
the result is independent of this choice.)  The evaluation of the Green
function part of the path integral (or operator evaluation) is then the
same as for $\cN=1$, since the physical $\cN=2$ variables are the same as those
for $\cN=1$ in $d=4$.  (However, unlike the $\cN=1$ case, spectral 
flow allows us to generalize to arbitrary boundary conditions: see 
below.)  The ghosts do not appear in the vertex operators,
and thus the Green functions, if we restrict the external states to be on
shell and gauge fixed.

Of course, the evaluation of the Green functions is not affected by the
anomaly, as they are the classical part of the JWKB expansion.  The
evaluation of the measure in the $\cN=2$ case is actually simpler.  As in the
$\cN=1$ and 2 cases, it depends on only the modulus ``$\tau$" related to the
sum of the Schwinger parameters:  Thus we can forget the external lines
and look at just the partition function.  We first examine the contribution
of the nonzero modes (oscillators).  In $d=4$ these contributions cancel
identically by counting:  The worldsheet variables of half-integral and
integral worldsheet spin (or conformal weight) each satisfy the same
boundary conditions, independent of statistics.  (This is true for arbitrary
choice of boundary conditions.)  But each of the two sets has equal
numbers of opposite statistics:  four $x$'s vs.\ $(b,c,b',c')$, and four
$\psi$'s vs.\ $(\beta,\gamma,\beta',\gamma')$.  Thus all the contributions
of the nonzero-modes cancel, leaving a partition function $f=1$.  (For
other dimensions, $f$ is a $d$-independent function to the power $d-4$.)

All these functional integration results so far agree with those 
found previously by invariance
arguments (see previous section).  The final step is the integration 
of the zero-modes.  The
integration over $x$ zero-modes is identical to that in ordinary field
theory, so we leave it for last, and do not discuss it.  That leaves only the
integration over the zero-modes of the fermionic ghosts.  Since $L_0$ has
no dependence on these zero-modes, these integrals are trivial:  If
$1/L_0$ were the complete propagator, they would give zero, since, e.g.,
$\int db_0 1 = 0$.  This problem is fixed in the same way as for $\cN=0$:  We
include in the full propagator a numerator of these zero-modes:
\bqr
\Delta' = {b_0 b'_0\over L_0}
\label{deltaprime}
\fqr
  In fact, this numerator is already required for evaluating tree graphs,
when ghosts are included.  The result is that the measure is identical to
that obtained by field theory methods.  (Earlier evaluations based on
invariance arguments did not directly address the problem of integration
over these zero-modes.)  The fact that it disagrees with the result
obtained from modular invariance arguments (though only by a factor of
$\tau$) is the anomaly in this invariance.

To expand on the above we derive the scattering amplitude for
a specified set of (unintegrated) spin structures of the worldsheet
fermions that specify an arbitrary spin.  This generalizes the 
calculation above, although in a trivial way because of the spin 
independence (spectral flow).  (Thus, in the supersymmetric theory 
the various contributions cancel.) The propagator discussed above is 
defined
as
\bqr
\langle X_1,\psi_1 \vert_\theta \int d\tau ~e^{-\tau L_0}
\vert X_2,\psi_2\rangle_{\theta}
= \langle X_1,\psi_1 \vert_\theta {1\over L_0} \vert X_2,\psi_2
\rangle_\theta \ ,
\fqr
with
\bqr
L_0 = {1\over 2} \sum_{n=-\infty}^\infty :a^\mu_{-n} a_{n,\mu} :
  + {1\over 2} \sum_{n=-\infty}^\infty (n+\theta) :
d^{\theta,\mu}_{-n}  d^\theta_{n,\mu} :
\ ,
\fqr
and
\bqr
d_n^{\theta} = d_{n+\theta} \quad d_{-n}^{\theta}=d_{-n-\theta} \quad 
 n>0 \ .
\label{dntwist}
\fqr
Here the oscillators are labeled by the twist angle $\theta$ that
interpolates between the Ramond and Neveu-Schwarz sectors, induced by the
spectral flow.  (There is a zero mode associated with the bosonic 
ghosts at $\theta=0$ that requires special treatment, which we do not 
discuss here.)
At every $\theta$ there is a complete set of states, and the
sectors are related by the spectral flow automorphism \rf{automorphism};
the spectral flow also twists the boundary conditions at pairs of
points where vertex operators are located, as described in section 3.

The tree amplitudes are constructed via
\bqr
A_n(k_i) = \langle k_1\vert^{\theta_1} \Bigl( {1\over L_0}\Bigr)
  V^{\theta_2}(k_2)
\ldots \Bigl({1\over L_0}\Bigr) \vert k_n\rangle^{\theta_n} \ ,
\label{tree}
\fqr
with twist parameters $\theta_i$ at each vertex which are required
to specify the boundary conditions of the worldspinors and thus the 
spin as described in
previous sections.  (We have not explicitly inserted the line factor
labeling the spin in this equation.)  The sewing relation
found by inserting a complete set of states in the loop generates
\bqr
A_n^{{\rm loop}} = \int {d^4p\over (2\pi)^4} {\rm Tr}_{\{\theta_i\}}
  \left( {1\over L_0} V_1^{\theta_1} {1\over L_0}
  V_2^{\theta_2} \ldots V_n^{\theta_n} \right) \ .
\label{sewedresult}
\fqr
We note that the factor inside the trace in
this equation is identical to the $\cN=1$ string amplitude evaluated
in four-dimensions (for $\theta$'s = 0 or 1/2), after normalizing the 
$\cN=2$ string vertex operators
with covariant line factors \rf{vertex}: The gauge-fixed action for the
matter components of the $\cN=1$ string has a global $\cN=2$ supersymmetry
on the world-sheet.

The complete expression following from the above is
\bqr
A_n (k_j) = \int_0^\infty {d\tau\over \tau^{3-n} } ~
  \int {\prod_{j=1}^n} dz_i ~ K_{KN}^{\theta_j}(z_i,\tau)  \ .
\label{untwistedamp}
\fqr
where the oscillators associated with the modes are inserted,
including the ghost terms but without the zero mode associated
with the latter.  The
integration generates in the zero-slope limit the integrand for the
maximally helicity violating amplitudes in four-dimensions; the
kinematic factor in \rf{untwistedamp} is
$$
K_{KN}^{\theta_j}(z_i,\tau_2) = \int \prod_{j=1}^n d^2\tilde\theta_j
\prod_{i<j} e^G
$$
\bqr
G = \Bigl[ -{1\over 2} k_i\cdot k_j G_{ij}
+ \epsilon_{[i}{\cdot}k_{j]} D^+_i G_{ij} +
\epsilon_i{\cdot}\epsilon_j D^+_i D^+_j G_{ij}
\Bigr]_{\rm multi-linear}
\label{bosoncont}
\fqr
where $D=\partial_{\tilde\theta^-}+\tilde\theta^- \partial_z$ and
$G_{ij}$ the two-point function between the points $(z_i,\tilde\theta^\pm_i)$
and $(z_j,\tilde\theta^\pm_j)$ for fermions of spin structure $\theta$.
The expansion in this equation is in products which are
linear in each polarization vector $\epsilon_j$.  This expansion may 
be found in \cite{Chalmers:2000wd} for general
choices of reference momenta of the polarization vectors 
$\epsilon_j^{++}$.  The complete integration over the modes in the 
$q$-expansion beyond the
massless sector is non-trivial, but the fermionic pieces at leading
order in $q$ integrate to zero by a supersymmetric Ward identity
\cite{Chalmers:2000wd}.

However, the contribution from the worldsheet Green functions is not the
same as from the field theory, although they agree in the zero-slope
limit.  This indicates that this anomaly is more serious, since contributions
from unphysical massive modes to the Green functions do not cancel, in
conflict with BRST invariance.  On the other hand, in the spacetime
supersymmetric case, contributions from spacetime fields of opposite
spacetime statistics cancel.  Thus, just as anomaly cancelation requires
SO(32) for the open superstring, anomaly cancelation for the $\cN=2$
strings requires spacetime supersymmetry.

At one loop, sewing can be replaced with a unitarity construction based
on dimensional continuation.  In the remainder of this section we examine
this  construction on the tree-amplitudes obtained through the field
theory describing  the classical dynamics. (The selfdual Lagrangians
reproduce the classical scattering  of the $\cN=2$ string.)  Then
two-particle unitarity cuts are sufficient  to determine the complete
amplitudes through an integral reduction of one-loop  integrals onto a
finite set of integral functions together  with their unique  cut
structures.
The unitarity cut in a two-particle channel containing momenta
$k_1+\ldots +k_m$ (with
$\sum_{j=1}^n k_j=0$) is
$$
\cI{\rm m} ~\cA_n^{[2]}(k_i) = \sum_{\lambda_i=\pm} \int d^dp d^dq ~
\delta^{(d)}(p^2) \delta^{(d)}(q^2) \Theta(p_0) \Theta(q_0)
\qquad\qquad\qquad
$$
\bq
\times \cA^\star_{m+2}(p^{\lambda_1},k_1,\ldots,k_m,q^{\lambda_2})
    \cA_{n-m+2}(q^{-\lambda_2},k_{n-m},\ldots,k_1,p^{-\lambda_1}) \ ,
\label{imaginary}
\fq
with helicity states denoted by $\lambda$
for an internal state of spin two. (The identical helicity configuration
generates the one-loop amplitude to the selfdual gravity theories, exact
in the two-field formulation.) Scattering of external states of identical
out-going helicity has tree amplitudes on either side of the unitarity cut
that equal zero: The tree amplitudes in four-dimensions with helicity
configuration $\cA(\pm,+,\ldots,+)$ equal zero at $n$-point. The absence
of unitarity cuts is manifest in the functional form \rf{allplusgauge};
continuation to $d=2+2$ dimensions results in the reality of the inner
products $\langle ij\rangle$ and $[ij]$ and the amplitude is then purely
real for all kinematic configurations.

Although the imaginary part of the amplitude in \rf{imaginary} is equal to
zero in four dimensions, upon continuing to $d=4-2\epsilon$ dimensions it
picks up a non-vanishing result (see, for example,
\cite{Bern:1997ja,Bern:1998xc}).
This unitarity construction leads to a direct  evaluation of the MHV
amplitude in different dimensions, and the integration  measure over the
cut loop momenta with the integration $d^dl$. However, this result is in
contradiction to the assumption of conformal invariance in the bosonic
$\cN=2$  string.

The sewing relation in string theory \rf{imaginary} precludes the moduli
associated with the $U(1)$ gauge field and the ghost determinant, i.e.,
the integration over the boundary conditions $(\alpha\beta)$ in
\rf{modmeasures}.  In the relation \rf{imaginary} we project onto a
finite number of massless states, found by summing discretely over the
spin structures.  The target space-time implementation in the sewing of
integrating out these fields (and the induced spectral flow) requires a
summation over intermediate massless states with spin ranging continuously
from zero to two.  In this construction of the loop the absence of
the $\tau_2$
factor from the $(b'c')$ ghost determinant in \rf{ghostdet} forces
the amplitude
to be evaluated in four-dimensions, and a choice of the spin structure in the
anti-periodic/anti-periodic, $\alpha=\beta=0$, sector maps the result to the
non-vanishing gravitational MHV amplitudes. (The periodic spin structures
generate a vanishing contribution holomorphically to the amplitude
\cite{Chalmers:2000wd} point-wise in the integration over $\tau$.)  Indeed,
the Koba-Nielsen representation of the amplitude from the $\cN=1$ string is
identical to that from the $\cN=2$ string without this factor; the $\cN=1$
string has a manifest global $\cN=2$ supersymmetry.  The sum over the
intermediate states in \rf{imaginary} with the appropriate supersymmetric
spectrum makes agreement with the direct path integral quantization including
the complete set of determinant factors in \rf{fulldetset}.

The tree amplitudes of the string theory agree with those of selfdual
theories: nonvanishing three-point amplitudes and vanishing
higher-point.  Sewing these trees, in either a field theory or
string description, straightforwardly produces vanishing one-loop amplitudes
for the space-time supersymmetric theories, but nonvanishing amplitudes
in the bosonic theories.  Since the assumption of modular invariance of
the one-loop amplitudes implies vanishing amplitudes, the one-loop
amplitudes violate conformal invariance unless the theory is spacetime
supersymmetric:  The one-loop amplitudes as derived by sewing are not
modular invariant in the nonsupersymmetric theories.

\section{Conclusions}

In this work we demonstrated the global conformal anomaly within  the
$\cN=2$ closed string.  This conformal
anomaly is potentially related to  an index in the $\cN=2$ string 
theory.  Cancelation of the anomaly requires, through
unitarity or sewing, that the theory be space-time supersymmetric.  The
supersymmetric extension of the $\cN=2$ closed string is accomplished
via the attachment of line factors on the vertex operators labeling spin
on the scalar state in the spectrum.  The bosonization  of states of different
spin in the field theory, or the spectral  flow of the $\cN=2$ system,
allows the incorporation  of supersymmetry through these factors.  A similar
analysis may be performed in the open string.

We further analyzed duality between open and closed string world-sheets
in the context of selfdual Yang-Mills theory and gravity and its
supersymmetric completion.  Duality in this context and in a theory
containing only a finite number of fields requires supersymmetry and
a vanishing S-matrix.  This is analyzed directly in the field theory
by showing that a general zero-momentum insertion into the maximally
helicity violating tree amplitude in Yang-Mills theory generates zero, in
accord with the supersymmetric closed selfdual gravity at the quantum
level to all orders in the loop expansion.

The fact that the nonvanishing $n$-point maximally helicity violating
loop amplitudes are due to such a simple anomaly suggests that an even
simpler derivation of these amplitudes might be possible, analogous to
the way the effective action of the Schwinger model follows simply from
its anomaly.  The Liouville multiplet offers an avenue to compensate the
factor of $\tau_2$ in the closed string calculation to obtain these
amplitudes.  Furthermore, the unintegrated zero momentum 
operator insertion noted in \cite{Chalmers:2000wd}, i.e., $\sqrt{g}~ 
\partial X^\mu{\bar\partial}  X_\mu$, representing the insertion of a 
tadpole would produce a factor of $\tau_2$ necessary to obtain these 
amplitudes.  A simple way to see this is to note that the one-loop 
amplitude can be written in terms of the trace of the Schwinger 
parametrized ($\tau_2$) propagator, and an extra $\tau_2$ corresponds 
to squaring the propagator,
\bq
tr(K^{-1}) = tr\left(\int_0^\infty d\tau\; e^{-\tau K}\right),\quad
tr(K^{-2}) = tr\left(\int_0^\infty d\tau\; \tau e^{-\tau K}\right) \ ,
\fq
which is the same as inserting a zero-momentum state between the
two propagators, i.e., a tadpole insertion.

\section*{Acknowledgments}

The work of G.C. is supported in part by the U.S. Department of Energy,
Division of High Energy Physics, Contract W-31-109-ENG-38, and that of
W.S. by NSF Grant PHY 9722101.

\vskip .3in 

\noindent{\it Note Added:}  The integration measure containing the spin 
structure summation that generates the following results is 
\bqr
{d^2\tau\over \tau_2^2}  \qquad\quad  {du d{\bar u}\over \tau_2^2}
  ={d\alpha d\beta\over\tau_2} \ .
\label{nonmodmeasures}
\fqr
It may be found by inserting into the {\it torus} amplitude (with the 
Jacobian torus associated with the gauge fields, or equivalently the spin 
structures, fibred on it) the ghost modes associated with the Beltrami 
differentials, $c{\bar c} b{\bar b} {\tilde c} {\tilde{\bar c}} {\tilde b} 
{\tilde {\bar b}}$; in this manner the zero modes of both the $(bc)$ and 
$(b'c')$ system are treated on an equal footing\footnote{We thank Olaf 
Lechtenfeld for a discussion on this aspect.}.


\end{document}